\documentclass[aps,prl,twocolumn,superscriptaddress,groupedaddress,showpacs]{revtex4}
\usepackage{graphicx,epsf}
\begin{document}
\title{ Where two fractals meet: the scaling of a self-avoiding
walk on a percolation cluster}
\author{C.\ von Ferber}
\email[]{ferber@physik.uni-freiburg.de} \affiliation{Theoretische
Polymerphysik, Universit\"at Freiburg,
                D-79104 Freiburg, Germany}
\affiliation{Institut f\"ur Theoretische Physik, Johannes Kepler
Universit\"at Linz, A-4040, Linz, Austria}
\author{V. Blavats'ka}\email[]{viktoria@icmp.lviv.ua}
\affiliation{Institute for Condensed Matter Physics, National
Academy of Sciences of Ukraine, UA--79011 Lviv, Ukraine}
\author{R. Folk}\email[]{folk@tphys.uni-linz.ac.at}
\affiliation{Institut f\"ur Theoretische Physik, Johannes Kepler
Universit\"at Linz, A-4040, Linz, Austria}
\author{Yu. Holovatch}\email[]{hol@icmp.lviv.ua}
\affiliation{Institut f\"ur Theoretische Physik, Johannes Kepler
Universit\"at Linz, A-4040, Linz, Austria}
 \affiliation{Institute for Condensed Matter Physics and
Ivan Franko National University of Lviv, UA--79011 Lviv, Ukraine}
\date{\today}
\begin{abstract}
The scaling properties of self-avoiding walks on a $d$-dimensional
diluted lattice at the percolation threshold are analyzed by a
field-theoretical renormalization group approach. To this end we
reconsider the model of Y. Meir and A. B. Harris (Phys. Rev.
Lett., {\bf 63}, 2819 (1989)) and argue that via renormalization
its multifractal properties are directly accessible. While the
former first order perturbation did not agree with the results of
other methods, we find that the asymptotic behavior of a
self-avoiding walk on the percolation cluster is governed by the
exponent $\nu_{\rm p}={1}/{2}+{\varepsilon}/{42}+
{110}\varepsilon^2/{21^3}$, $\varepsilon=6-d$. This analytic
result gives an accurate numeric description of the available MC
and exact enumeration data in a wide range of dimensions $2\leq d
\leq 6$.
\end{abstract}
\maketitle

Polymers and percolation clusters are among the most frequently
encountered examples of fractals in condensed matter physics
\cite{deGennes79,Stauffer,Mandelbrot83}. When a long polymer chain
is immersed in a good solvent its mean-square end-to-end distance
$\overline{R^2}$ scales with the monomer number $N$ as:
\begin{equation} \label{1}
\overline{R^2} \sim  N^{2 \nu_{\rm {SAW}}}, \hspace{3em}
N\rightarrow\infty
\end{equation}
with the exponent $\nu_{\rm SAW}(d)$ which depends on the
(Euclidean) space dimension $d$ only. This scaling of polymers
(\ref{1}) is perfectly described by the self-avoiding walk (SAW)
on a {\em regular} $d$-dimensional lattice \cite{deGennes79} and
the fractal dimension of a polymer chain readily follows: $d_{\rm
SAW}=1/\nu_{\rm SAW}$. For space dimensions $d$ above the upper
critical dimension $d_{\rm up}=4$ the scaling exponent becomes
trivial: $\nu_{\rm SAW}(d> 4)=1/2$, whereas for $d<d_{\rm up}$ the
non-trivial dependence on $d$ is described e.g. by the
phenomenological Flory formula \cite{deGennes79} $\nu_{\rm
SAW}=3/(d+2)$. This found its further support by the
renormalization group (RG) $\tilde\varepsilon=4-d$-expansion known
currently to the high orders \cite{Kleinert}: $\nu_{\rm
SAW}=1/2+\tilde\varepsilon/16+15\,\tilde\varepsilon^2/512+\dots$.

When a SAW resides on a {\em disordered} (quenched diluted)
lattice -- such a situation might be experimentally realized
studying a polymer solution in a porous medium, but is of its own
interest as well -- the asymptotic scaling behavior is a more
subtle matter \cite{theory,Kim83,Barat95}. Numerous MC simulations
\cite{Kremer81,Lee88,Lee89,Woo91,Grassberger93,Lee96} and exact
enumeration studies
\cite{Meir89,Lam90,Nakanishi92,Vanderzande92,Rintoul94,Ordemann00,Nakanishi91},
which last since early 80-ies \cite{Barat95}, lead to the
conclusion that there are the following regimes for the scaling of
a SAW on a disordered lattice: (i) weak disorder, when the
concentration $p$ of bonds allowed for the random walker is higher
than the percolation concentration $p_{\rm PC}$ and (ii) strong
disorder, directly at $p=p_{\rm PC}$. By further diluting the
lattice to $p<p_{\rm PC}$ no macroscopically connected cluster,
``percolation cluster'', remains and the lattice becomes
disconnected. In regime (i) the scaling law (\ref{1}) is valid
with the same exponent $\nu_{\rm SAW}$ for the diluted lattice
independent of $p$, whereas in case (ii) the scaling law (\ref{1})
holds with a new exponent $\nu_{\rm p} \neq \nu_{\rm SAW}$. A hint
to the physical understanding of these phenomena is given by the
fact that weak disorder does not change the dimension of a lattice
visited by a random walker, whereas the percolation cluster itself
is a fractal with fractal dimension dependent on $d$: $d_{\rm
PC}(d)=d-\beta_{\rm PC}/\nu_{\rm PC}$, where $\beta_{\rm PC}$ and
$\nu_{\rm PC}$ are familiar percolation exponents \cite{Stauffer}.
In this way, $\nu_{\rm SAW}(d)$ must change along with the
dimension $d_{\rm PC}$ of the (fractal) lattice on which the walk
resides. A modified Flory formula \cite{Kremer81} for the exponent
of a SAW on the percolation cluster $\nu_{\rm p}=3/(d_{\rm PC}+2)$
along with results of similar theoretical studies
\cite{Sahimi84,Rammal84,Kim87,Roy87,Aharony89,Kim90,Roy90,Kim92}
gives numbers in an astonishing agreement with the data observed
(see Table \ref{tab1}). Since $d_{\rm up}=6$ for percolation
\cite{Stauffer}, the exponent $\nu_{\rm p}(d\geq 6)=1/2$
\cite{Janssen03}.

Although the Flory-like theories
\cite{Sahimi84,Rammal84,Kim87,Roy87,Aharony89,Kim90,Roy90,Kim92}
offer good approximations for $\nu_{\rm p}(d)$ in a wide range of
$d$, even more astonishing is the fact that up to now there do not
exist any satisfactory theoretical estimates for $\nu_{\rm p}(d)$
based on a more refined theory, which takes into account
non-Markovian properties of the SAW, a task which was completed
for regular lattices already in mid-70-ies \cite{deGennes79}.
Existing real-space RG studies \cite{Roy82,Lam84,Sahimi84,Meir89}
give satisfactory estimates for $d=2$, whereas the
field-theoretical approaches aimed to describe the situation at
higher dimensions lead to contradictory conclusions. In
particular, the field theory developed in Ref. \cite{Meir89}
supported $d_{\rm up}=6$ and presented a calculation of $\nu_{\rm
p}$ in the first order of $\varepsilon=6-d$. However the numerical
estimates obtained from this result are in poor agreement with
numbers observed by other means, leading in particular to the
surprising estimate $\nu_{\rm p}\simeq\nu_{\rm SAW}$ in $d=3$ (see
Table \ref{tab1}). In turn, a subsequent study \cite{Ledoussal91}
even questioned the renormalizability of this field theory and
suggested another theory with $d_{\rm up}=4$ which is obviously
disproved by computer simulations and exact enumerations at
dimensions $d=4,5$ \cite{Lee96,Meir89}.

\begin{table}
\begin{small}
\begin{tabular}{|r| c| c|  c|  c|  c| }
\hline $d$  & 2 & 3 & 4& 5 & 6 \\ \hline $\nu_{\rm SAW}$ & 3/4
 & 0.5882(11)  &  1/2
&1/2   &1/2  \\ \hline FL, \cite{Sahimi84}
  &
0.778& 0.662 & 0.593 & 0.543 & 1/2\\ \cite{Rammal84} & 0.69(1)&
0.57(2) & 0.49(3) &  & 1/2\\
 \cite{Kim87} & & 0.70(3)&
0.63& 0.56&1/2\\ \cite{Roy87}  & 0.770& 0.656 & 0.57 & 0.52 &
1/2\\
 \cite{Aharony89} & 0.76&
0.65 & 0.58 &  & 1/2\\
 \cite{Kim90} & 0.75-0.76& 0.64-0.66 &
 0.57-0.59 & 0.55-0.57 & 1/2\\
\cite{Roy90} & 0.77& 0.66 &
 0.62 & 0.56 & 1/2\\
  \hline MC, \cite{Kremer81} & &
$\simeq$ 2/3 & & &
\\
\cite{Lee88}
  & $\simeq \nu_{\rm SAW}$  & 0.612(10) &  & & \\
\cite{Lee89}
  & $\simeq \nu_{\rm SAW}$  & 0.605(10) &  & & \\
\cite{Woo91}
  & 0.77(1)  & &  & & \\
\cite{Grassberger93} & 0.783(3) & & & &
\\
\cite{Lee96} & & 0.62-0.63 &0.56-0.57 & &
\\
 \hline EE, \cite{Meir89}  &
0.76(8)
 &0.67(4) & 0.63(2) & 0.54(2) & \\
 \cite{Lam90}
&0.81(3)& &&& \\ \cite{Lam90} &0.745(10)& 0.635(10)&&&\\
\cite{Nakanishi92}&& 0.65(1)&&& \\ \cite{Vanderzande92}&0.745(20)&
0.640(15)&&&\\ \cite{Rintoul94}&0.770(5)& 0.660(5)&&&\\
\cite{Ordemann00}&0.778(15)& 0.66(1)&&&\\
\cite{Ordemann00}&0.787(10)& 0.662(6)&&&\\
 \hline RS,
\cite{Lam84} & 0.767 & &  & & \\ \cite{Sahimi84} & 0.778&0.724
&&&\\ \hline RG, \cite{Meir89} & 0.595 & 0.571& 0.548 & 0.524 &1/2
\\ (\ref{8}) & 0.785 & 0.678& 0.595 & 0.536 &1/2 \\
 \hline
\end{tabular}\end{small}
\caption{ \label{tab1} The exponent $\nu_{\rm p}$ for a SAW on a
percolation cluster. FL: Flory-like theories, EE: exact
enumerations, RS, RG: real-space and field-theoretic RG. The first
line shows $\nu_{\rm SAW}$ for SAW on the regular lattice ($d=2$
\cite{Nienhuis82}, $d=3$ \cite{Guida98}).}
\end{table}

There is another important reason, why the scaling of a SAW on a
percolation cluster calls for further theoretical study. As it
became clear now, higher-order correlations of a fractal object at
another fractal lead to multifractality \cite{MF}. Recently
studied examples of multifractal phenomena are found in such
different fields as diffusion in the vicinity of an absorbing
polymer \cite{absorber}, random resistor networks \cite{resistor},
quantum gravity \cite{Duplantier99}. A SAW on a percolation
cluster is a good candidate to possess multifractal behavior.
Indeed such behavior is found in computer simulations
\cite{Ordemann00}, moreover it naturally emerges in the RG scheme,
as we will explain below.

Let us consider a diluted lattice with sites ${\bf x}_i$ in terms
of variables $p_{ij}=0,1$ that indicate whether a given bond
between the sites ${\bf x}_i$ and ${\bf x}_j$ is present or not.
To describe the critical properties of SAWs on this lattice
following the idea of de Gennes \cite{deGennes79} we introduce
$m$-component spin variables $S_{\alpha}({\bf x}_i)$,
$\alpha=1,\dots,m$, and evaluate the theory for $m=0$. To allow
for the averaging over the {\em quenched} disorder the spins are
$n$-fold replicated which gives for the Hamiltonian:
\begin{equation}\label{2}
{\rm e}^{-{\cal H}_S}=<\exp\{-\frac{K}{2}\sum_{i,j}
p_{ij}\sum_{\alpha=1}^m \sum_{\beta=1}^n S_{\alpha}^{\beta}({\bf
x}_i)S_{\alpha}^{\beta}({\bf x}_j \}>_p
\end{equation}
where we denote by $<\dots>_{p}$ the average over the random
variables $p_{ij}$ which take the value 1 and 0 with probabilities
$p$ and $(1-p)$ respectively, and $K$ is an interaction parameter.
In the following we will work with a field theoretical
representation of the effective Hamiltonian defined in (\ref{2}).
This is achieved \cite{Meir89} via a Stratonovich-Hubbard
transformation to tensor fields $\psi_{k}({\bf x })$ with
components
$\psi_{k;\beta_1,\dots,\beta_k}^{\alpha_1,\dots,\alpha_k}({\bf x
})$ conjugated to the product $\Pi_{j=1}^k
S_{\alpha_j}^{\beta_j}({\bf x})$ of $k$ components of the
replicated spin with $\beta_1<\dots<\beta_k$. This results in the
effective Hamiltonian up to order $\psi^3$
 \cite{Meir89}:
 \begin{eqnarray}\nonumber
{\cal H}_{\psi} &=& \frac{1}{2}\int {\rm d}^d q \sum_k
 (r_k+q^2)\psi_k({\bf q}):\psi_k(-{\bf q}) +
\\ && \label{3}
 \frac{w}{6}
 \int {\rm d}^d x \psi^3({\bf x}),
 \end{eqnarray}
where $\psi_k({\bf q})$ is the Fourier transform of $\psi_k({\bf
x})$, the inner product reads: $$ \psi_k({\bf q}):\psi_k(-{\bf
q})= \sum_{\{\alpha_i\}} \sum_{\{\beta_i\}}
|\psi_{k;\beta_1,\dots,\beta_k}^{\alpha_1,\dots,\alpha_k}({\bf
q})|^2 ,$$ and $\psi^3({\bf x})$ is a symbolic notation for a
product of three $\psi_k$ fields. Only those cubic terms $\psi^3$
are allowed for which all pairs $(\alpha_i,\beta_i)$ appear
exactly twice. A second condition on the diagrammatic
contributions to perturbation theory can be derived from the de
Gennes limit $m=0$, namely, if any index $(\alpha,\beta)$ appears
only on the internal propagator of a diagram, then its
contribution vanishes.

We note the unusual dependence of ``masses" $r_k$ on $k$. This is
reminiscent of the fact that in the $m=0$ limit the theory
(\ref{2}) becomes multicritical \cite{Meir89,Derrida84}. This has
impact on the renormalization of the theory (\ref{3}) as we will
show in the following.

\begin{figure} [!htb]
\centerline{\includegraphics[width=70mm]{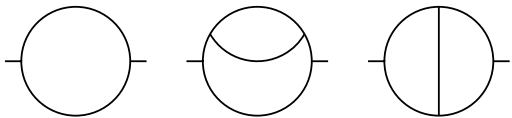}}
\caption{\label{fig1} The Feynman graphs of the vertex function
$\Gamma^{(2)}(q)$ in the two lowest orders.}
\end{figure}

We choose to calculate the critical properties of the theory by
analyzing its vertex functions, in particular $\Gamma^{(2)}(q)$,
$\Gamma^{(3)}(\{q\})$, and $\Gamma^{(2,1)}(\{q\})$ where the
latter includes an insertion of the $\psi\mbox{:}\psi$ operator.
Each of these $\Gamma$-functions will depend on the family of
masses $\{r_k\}$. The Feynman graphs of the contributions to the
two-point vertex function $\Gamma^{(2)}(q)$ in the two lowest
orders are shown in Fig. \ref{fig1}.
The contributions to $\Gamma^{(2,1)}$ are found from this by
placing an insertion on each of the inner propagator lines. These
integrals are evaluated then in dimensional regularization in
dimension $d=6-\varepsilon$ and minimal subtraction
\cite{tHooft72} using a Laurent-expansion in $\varepsilon$.
Usually the renormalization of the vertex functions is defined in
terms of $Z$-factors in such a way that the products
$Z_{\psi}\Gamma^{(2)}$, $Z_w\Gamma^{(3)}$,
$Z_{\psi^2}\Gamma^{(2,1)}$ are free of $\varepsilon$-poles.
However, the insertion of the $\psi\mbox{:}\psi$-operator together
with the $k$-dependence of the masses $r_k$ leads to the following
renormalization procedure. The vertex function $\Gamma^{(2,1)}$
even when evaluated at zero mass remains $k$-dependent:
\begin{equation}\label{4}
\Gamma^{(2,1)}=\Gamma^{(2,1)}_0 + k \Gamma^{(2,1)}_1 + k^2
\Gamma^{(2,1)}_2+\dots
\end{equation}
and it can not be renormalized by one multiplicative $Z$-factor.
The essential feature of this expansion is that each term shows a
different scaling behavior. In this way the multicriticality
recognized already by Derrida \cite{Derrida84} and Meir and Harris
\cite{Meir89} manifests itself in our present formalism and leads
to a {\em spectrum of exponents}. Instead of a single $Z$-factor
$Z_{\psi^2}$ a whole family of factors $Z_{\psi:\psi}^{(i)}$ is
necessary  to renormalize each $\Gamma^{(2,1)}_i$ in (\ref{4}).
This allows to define RG-functions
\begin{equation}\label{5}
 \beta(w)= \frac{\rm d}{{\rm d} \kappa} \ln Z_w,
\end{equation}
that describes the RG-flow of the coupling with respect to the
rescaling parameter $\kappa$ and
\begin{equation}\label{5a}
 \eta(w)= \frac{\rm d}{{\rm d} \kappa} \ln Z_{\psi},
 \qquad
 \eta_{\psi:\psi}^{(i)}(w)=
  \frac{\rm d}{{\rm d} \kappa} \ln Z_{\psi:\psi}^{(i)}
\end{equation}
which define the anomalous dimensions of the corresponding
operators. At the stable fixed point $w^*$ with $\beta(w^*)=0$ the
family of correlation exponents is given by
\begin{equation}\label{6}
\nu^{(i)}= [2-\eta(w^*)+ \eta_{\psi:\psi}^{(i)}(w^*)]^{-1}.
\end{equation}
We note that $\nu^{(0)}\equiv\nu_{\rm PC}$ and
$\nu^{(1)}\equiv\nu_{\rm p}$ as introduced above, whereas the
$\nu^{(i)}$ for $i\geq 2$ are connected with higher order
correlations. The $\beta$-function (\ref{5}) is the familiar RG
function of the $\psi^3$ Potts model \cite{Amit76}.

The explicit calculations proceed as follows: (i) One starts with
the vertex function $\Gamma^{(2)}_{\psi_k}$ corresponding to the
propagator of the field $\psi_k$. (ii) For the masses one inserts
the expansion $r_k=\mu\sum_{j=0}^\infty u_jk^j$. (iii) The
insertion of $\psi\mbox{:}\psi$ is defined by the derivative
$\frac{\partial}{\partial \mu} \Gamma^{(2)}_{\psi_k}$ evaluated at
zero mass for $\mu=0$. (iv) Performing the summation over the
replica indices the contributions to the different
$\Gamma^{(2,1)}_i$ are generated by rearranging the expansion in
$k$. One finds the multiplicative renormalization for
$\Gamma^{(2,1)}_i$ for appropriate linear combinations of the
different orders of $k$.

\begin{figure} [!htb]
\centerline{\includegraphics[width=70mm]{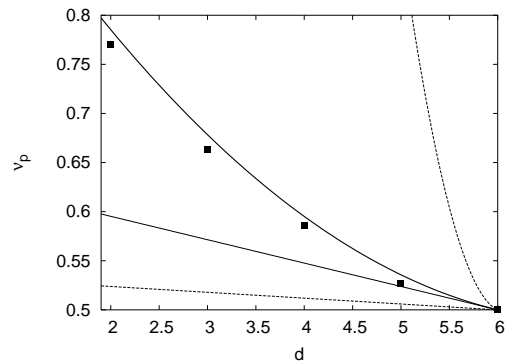}}
\caption{\label{fig2} The correlation exponent $\nu_{\rm p }$.
Bold line: (\ref{8}), thin line: one-loop result \cite{Meir89},
filled boxes: Flory result $\nu_{\rm p}=3/(d_{\rm PC}+2)$ with
$d_{\rm PC}$ from \cite{Df}. Exponents for the shortest and
longest SAW on percolation cluster \protect\cite{Janssen} are
shown by dotted lines.}
\end{figure}

Following this procedure we obtain $\varepsilon=6-d$ expansions
for  $\eta_{\psi:\psi}^{(i)}$. Substituting them together with the
known result \cite{Amit76} $\eta=-\varepsilon/21 -
206\varepsilon^2/21^3$ into (\ref{6}) we arrive at the following
spectrum of correlation exponents:
\begin{eqnarray}\label{7}
\nu^{(0)}&=&\nu_{\rm PC}=1/2 + 5\, \varepsilon/84 + 589\,
\varepsilon^2/42^3,\\ \label{8}
 \nu^{(1)}&=&\nu_{\rm p}=1/2 +
\varepsilon/42 + 110\, \varepsilon^2/21^3,\\ \label{9}
\nu^{(2)}&=&1/2 + \varepsilon/24 + 13907\, \varepsilon^2/1100736,
\dots.
\end{eqnarray}
By (\ref{7}) we recover the familiar $\varepsilon$-expansion for
the percolation exponent $\nu_{\rm PC}$ \cite{Amit76} and in
(\ref{8}) we extend the first order result for $\nu_{\rm p}$
\cite{Meir89}. The physical interpretation and properties of the
remaining exponents $\nu^{(i)}$ of the family is the subject of a
separate study \cite{unpublished}. Contrary to the family of
$\nu$-exponents defined in Ref. \cite{Ordemann00}, the $\nu^{(i)}$
govern the non-trivial scaling of properly defined cumulants of
the distribution of SAWs for given end-to-end distance
\cite{unpublished}. Evaluating the result for $\nu_{\rm p}$
(\ref{8}) by direct substitution of $\varepsilon=6-d$ one finds
nearly perfect correspondence with available MC and exact
enumeration results over the range $d=2,\dots,5$, see Table
\ref{tab1}. This presents a {\em qualitative} improvement over the
linear result as seen in Fig. \ref{fig2} where we also show that
the result  is in between the limits given by the shortest and
longest SAWs on percolation cluster \cite{Janssen}.

A rather peculiar finding is that results of the phenomenological
Flory-like formulae evaluated using the fractal characteristics of
the percolation cluster are numerically very close to our result
in the same region of dimensions. Note however the ambiguity
\cite{Rintoul94} in defining a Flory-like scheme leading to the
different results in Table \ref{tab1}.

The $\psi^3$ theory as applied to the present problem inevitably
has the upper critical dimension $d_{\rm up}=6$. This in
particular allows us to describe the discussed non-trivial scaling
for dimensions $d=4,5$. This is out of reach following the
approach of Ref. \cite{Ledoussal91} which gives trivial scaling
for $d\geq d_{\rm up}=4$ and relies on a $\phi^4$-theory with two
couplings of different symmetry. Moreover, in the de Gennes limit
$m=0$ the symmetry of the two couplings  coincides \cite{Kim83}
leading back to a theory of SAWs on the pure lattice with a
redefined coupling parameter, a fact neither exploited in Ref.
\cite{Ledoussal91} nor in the similar approaches
\cite{Chakrabarti81,Obukhov90}.

From the physical point of view, our result for the exponent
$\nu_{\rm p}$ together with the data of EE and Flory-like theories
(see Table \ref{tab1}) predicts a swelling of a polymer coil on
the percolation cluster with respect to the pure lattice:
$\nu_{\rm p}>\nu_{\rm SAW}$ for $d=2-5$. Up to now, this
phenomenon has clearly been observed only in MC simulations for
$d=2$ \cite{Grassberger93}. Although simulations on $d=3$
percolation clusters have been claimed to show this effect
\cite{Kremer81,Lee88,Lee89,Lee96}, these studies were subsequently
criticized for using inappropriate data analysis
\cite{Lee88,Lam90,Nakanishi91} and for lack of accuracy. At $d=3$
our formula (\ref{9}) predicts a 13\% increase of $\nu_{\rm p}$
with respect to $\nu_{\rm SAW}$ which is larger than at $d=2$
(5\%) and should be more easily observed by current state-of-art
simulations. Given that even at $d=2$ we are in  nice agreement
with MC and EE data and the reliability of the perturbative RG
results increases with $d$, this number calls for verification in
MC experiments of similar accuracy.

\section*{Acknowledgements}
We thank Guy Bonneau, Bertrand Delamotte, and Verena
Schulte-Frohlinde  for enlightening comments. R.F. and Yu.H. thank
the Austrian Fonds zur F\"orderung der wissenschaftlichen
Forschung, project No.16574-PHY which supported in part this
research.

\end{document}